# DNA Replication Timing, Genome Stability and Non-adaptive Radiation

John Herrick, independent researcher
3, rue des Jeûneurs
Paris, France
 jhenryherrick@yahoo.fr

Abstract

A correlation between karyotype diversity and species richness was first observed in mammals in 1980, and subsequently confirmed after controlling for phylogenetic signal. The correlation was attributed to "submicroscopic factors", presumably operating at the level of the genome. At the same time, an unexpected association between mutation rates and substitution rates has been observed in all eukaryotes so far examined. One hypothesis to explain the latter observation proposed that neutral mutation (dS) and non-neutral (dN) substitution rates in gene codons co-vary according to genomic position, or location in the genome. Later, it was found that mutation and substitution rates in eukaryotes increase with DNA replication timing during the Synthetic phase, or S phase, of the cell cycle. In 1991, Motoo Kimura proposed a molecular theory of non-adaptive radiation (NAR). Accordingly, genetic drift plays a significant role in speciation, albeit in parallel to and in conjunction with natural selection. The following will examine the contribution of DNA replication timing and DNA repair factors to the relationship between species richness and karyotype diversity, and by extension, to the large variation in species richness across the mammalian Tree of Life.

Significance

The eukaryote DNA replication timing program (RT) organizes the DNA synthetic phase (S phase) of the cell cycle and coordinates genome duplication with mitosis and cell division. A complex system of DNA damage detection and repair (DDR) reinforces this organization in order to sustain and constrain mutation/substitution balance. The RT therefore has important implications for the evolution of genome architecture, karyotype diversity and species richness. To date, few studies have directly examined the role RT plays in speciation and adaptive radiations.

1. Introduction

1.1 Bengtsson's Hypothesis

Species richness, or the number of species in a taxonomic clade, varies widely across the Class Mammalia. At the Family taxonomic level, for example, the Muridae harbour 876 species out of 6,759 documented mammalian species, while some 24 other Families out of 167 contain only one extant species. At the Order level, mammalian diversity is also highly uneven: Rodentia and Chioptera account for 60 % of species while the other 25 Orders account for 40 % (1) The origin and causes of this imbalance (species evenness) are topics of longstanding interest and currently the focus of intense phylogenetic and phylogenomic investigation.

In 1980, Bengtsson discovered an unexpected correlation between species richness and karyotype diversity across the Class Mammalia (2). The striking correlation could be explained either by differential selection acting on adaptive phenotypes and life history traits, or by genetic drift promoting the survival or the extinction of randomly emerging traits in isolated species having small effective population sizes ($Ne$). Bengtsson proposed, in contrast, that most if not all of the karyotype diversity observed across the Mammalia clades might be attributable to submicroscopic, and hence molecular, factors. He did not speculate, however, what might be the factors in the cell or nucleus that contribute to variations species richness (SR).

Bengtsson states: "No one has yet demonstrated or suggested any karyotypic feature which in mammals characterizes relatively stable karyotypes. This failure to find properties characteristic of stable or unstable karyotypes may indicate that the cytological factors of importance for karyotype evolution are all of a sub-microscopic nature."

1.2 Karyotype evolution

Earlier findings in the 1970's revealed that karyotypes in Salamander, Mammalian and Angiosperm groups are evolving at significantly faster rates than the respective genes. Maxson and Wilson, for example, reported that amphibian karyotypes evolve at several times the rate of the genes residing in the corresponding genomes; a similar observation had also been made in plants (3, 4, 5). They calculated that evolutionary changes in the karyotype of salamanders are at least 20 times slower than in mammals and two times slower than anurans.

Moreover, earlier studies had suggested that evolution at the organism level is correlated more highly with karyotypic evolution than with structural gene evolution (3, 6, 7, 8). Levin and Wilson, for example, observed a striking correlation between rates of speciation and rates of karyotype diversification in plants (5): small herbaceous plants evolve faster and have correspondingly higher rates of karyotype change compared to species with larger biomasses (BM). Karyotypes, however, are presumably under relaxed selection and evolve more randomly (punctually) according to genetic drift, while genes are more subject to purifying and positive selection and evolve at a more constant rate. Consequently, orthologous genes typically diverge more slowly than the karyotypes of the respective genomes.

Additionally, vertebrates and invertebrates have approximately the same number of genes: between 14,000 and 25,000, the so-called "G-paradox" (9), and evolve on average according to a relatively constant molecular clock (10, 11). In contrast, rates of karyotype evolution, measured as rates of macro-karyotype changes (rKD Macro) or micro-karyotype changes (rKD Micro), vary widely across the mammalian phylogenetic clades and, by extension, across the Tree of Life (12). This observation is somewhat puzzling given that gene order, or synteny, is highly conserved in Mammalia, Aves and Amphibia (13, 14), indicating that synteny is under strong purifying selection (13).

Assuming that macroevolution (species diversification) is linked to microevolution (genetic divergence), the conserved average rate of mutations in protein coding exons suggests that a relatively constant (but varying molecular clock of 2- 4 fold) (10, 11) governs the rate of gene and genotype evolution (15, 16, 17), and therefore the rate at which species diverge (speciation rate) (18, 19). The hypothesis of speciation rate being constrained by a universal molecular clock remains, however, to be fully confirmed (20, 21).

1.3. Kimura's Theory of Non-adaptive Radiation

The apparent neutral evolution of KD and related expansions and contractions in neutral non-coding DNA are believed to result in a wide range of genome size in any given phylogenetic lineage (Figure 1). The stochastic expansion or contraction in a lineage's genome size (22, 23), such as the salamander lineage (24), led to the proposal of a second "junk" DNA clock (25). It should be noted, however, that the rate junk DNA changes in the genome is comparable to the respective mutation rate. In Drosophila, for example, the rate of transposon element (TE) activity is roughly similar to the mutation rate (ref).

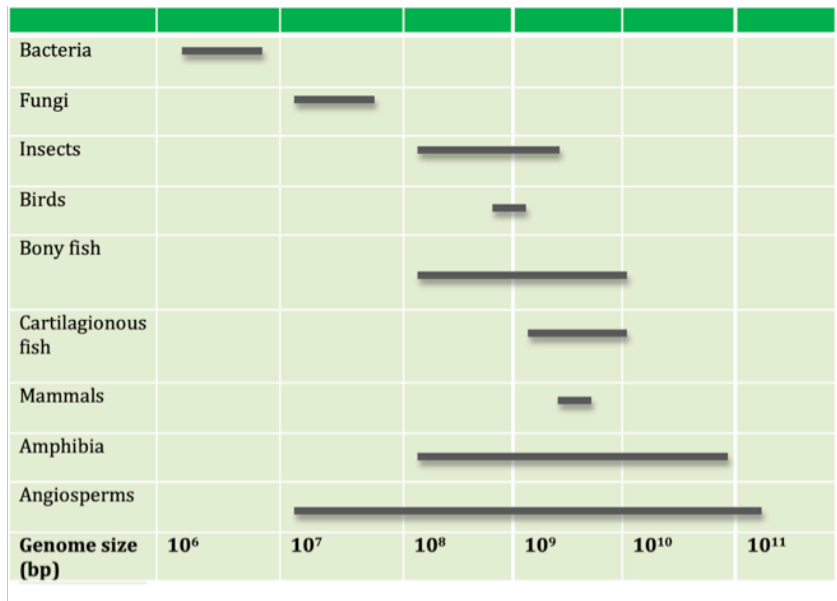

**Figure 1.** Schematic illustration (not to scale) depicting approximate ranges in genome size in different taxa. Amphibia and angiosperms exhibit a broad range in genome size compared to birds and mammals, resulting from large-scale DNA deletions and amplifications. Under genetic drift, genome size expansion is considered predominantly selectively neutral or nearly neutral (130). The degree to which selection and genetic drift shape the distribution of genome sizes, however, is an unsettled issue (131, 132).

The following sequence of events might properly frame the process of ecological succession that characterizes macroevolution (26; Figure 2):

1) geographic isolation following a population split.
2) neutral (non-genic) karyotype diversification driven by genetic drift, eventually involving genes in species with small effective population sizes (microevolution).
3) reproductive (pre and post-zygotic) isolation separating diverged populations (for example, ring species).
4) selection driving speciation and adaptive radiation into newly evolved niches (macroevolution).

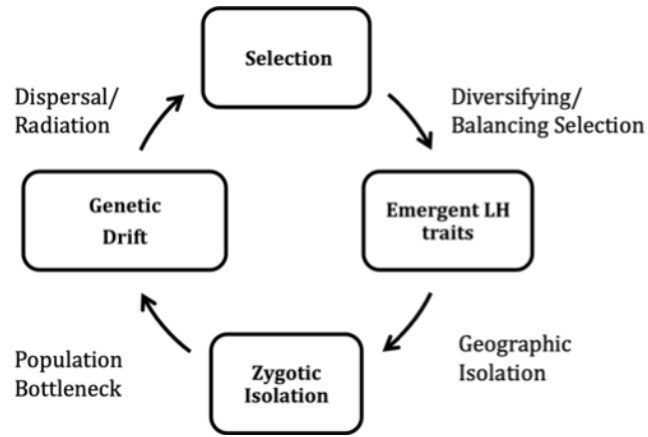

**Figure 2.** Generalized model of Non-adaptive Radiation (from Kimura 1991). Darwinian selection/speciation is believed to depend on the standing level of allelic diversity in a population (adaptation/diversification). Smaller effective population sizes are expected to have higher levels of mutation accumulation ("Drift Barrier Hypothesis"; see 133) The figure proposes that genetic drift (GD) can promote elevated mutation levels that can likewise result in increased levels of allelic diversity (population bottle neck). The interplay between these two forces, genetic drift and selection, determines the extent to which a phenotype or population primarily experiences an adaptive radiation or a non-adaptive radiation. In the absence of strong purifying selection, lineages with high karyotype diversity have a greater probability of being relatively species rich, while species with low karyotype diversity experiencing strong selection can have either low or high levels of species richness. Birds and frogs, for example, have conserved karyotypes but are species rich, whereas salamanders have low karyotype diversity and relatively low levels of species richness. In mammals, karyotype diversity explains about 86 % of variation in species richness at the mammalian Order level and 43 % at the Family level (Table 1).

|  | *Adj $R^2$* | *P* |
|---|---:|---:|
| **MLS vs Body Mass (+)** | **0.8** | **$6 \times 10^{-4}$** |
| *MLS vs C-value (+)* | 0.007 | 0.5 |
| *MLS vs KD (-)* | 0.29 | 0.08 |
| ***MLS vs Synteny (+)*** | **0.48** | **0.03** |
| ***SR vs Body Mass (-)*** | **0.56** | **0.01** |
| ***SR vs MLS (-)*** | **0.59** | **0.016** |
| ***SR vs KD (+)*** | **0.86** | **$7.2 \times 10^{-6}$** |
| ***SR vs rKD Macro (+)*** | **0.85** | **$1 \times 10^{-5}$** |
| *SR vs rKD Micro (+)* | 0.07 | 0.06 |
| *SR vs Synteny (-)* | 0.18 | 0.1 |

**Table 1.** Order level phylogenetic Generalized Least Squares (PGLS) correlations between mammalian life history traits. MLS: maximum lifespan. SR: species richness. rKD Macro: rate of karyotype evolution (genome wide structural changes: number of chromosomes and number of chromosome arms). rKD Macro reflects the level of karyotype diversity with respect to species richness ($R^2 = 0.86$; *p = $7.2 \times 10^{-6}$). rKD Micro:

rate of sub-chromosomal changes that do not alter the number of chromosomes or the number of chromosome arms. The rKD Macro and rKD Micro data are from Martinez *et al* (12).

2. Mutation Substitution Balance

2.1 Correlations between Neutral (dS) and Non-neutral Substitutions (dN)

Amino acid substitution rates in proteins have been shown to be proportional to nucleotide substitution rates in genes: non-synonymous (amino acid changing) substitutions in codons correlate with synonymous (silent) substitutions in genes in all eukaryotes examined (27, 28, 29, 30) The ratio between these two is therefore relatively constant (dN/dS proportional to 1). Any detectable deviations from neutrality (dS = dN) are interpreted as signatures either of purifying selection (dN/dS << 1) or positive selection (dN/dS > 1).

The correlation between dN and dS, though lacking a comprehensive molecular explanation, is nonetheless expected if, as commonly assumed, most amino acid substitutions in proteins are either deleterious or functionally neutral (dN and dS both reflect the mutation rate). The correlation, however, is much stronger than would be expected assuming that dN/dS simply reflects the proportion of neutral non-synonymous substitutions, and therefore the underlying mutation rate (30, 31). Moreover, the effect is uniform across all genes in a genome and phylogenetically independent of species relatedness (32, 33).

Several hypotheses have been advanced to explain the correlation (see for example: 30, 33). One hypothesis proposes that the correlation is due to a positional, or genomic context effect: because the mutation rate varies across the genome from yeast to plants and animals, any positional, or regional, change in the mutation rate will impact both dN and dS similarly (34). Both categories of mutation will be affected in equal proportion: region-wide dN and dS rates will increase or decrease together independently of the fact that different genes in the same genome experience significantly different mutation rates (32).

The effect, the hypothesis predicts, will also apply to non-genic, non-coding DNA residing in the same region as coding DNA. Polymorphisms, for example, in ultra-conserved elements (UCE), which are involved in vertebrate development and reside within introns or outside genes (35), correlate with dN and dS inside the exons and codons of the adjacent genes (27). The two categories of mutation, dN and dS, therefore remain correlated with polymorphisms in non-coding inter-genic regions, and perhaps intra-genic introns (36, 37), in support of the proposal that dN and dS are correlated locally within defined regions of the chromosome and genome.

2.2 Balancing Selection and Locus versus Gene Specific Mutation Rates

Analogous to the clonal selection theory in immunology, balancing selection on DNA polymorphisms and allelic diversity has acted to multiply the adaptive opportunities and evolutionary trajectories that have led to the emergence of increasingly complex organisms. This has been shown to be the case for certain genes in the immune system-related major histo-compatibility complex (MHC). The MHC replicates in the first half of S phase; but the class II elements (AT rich), compared to classes I and III (GC rich), replicate later at the middle of S phase (32).

Another study also revealed a correlation between dN and dS but instead with dN > dS, indicating selection for diversity. In salamanders, for example, levels of dN/dS are significantly higher than in other vertebrates while levels of dS are, paradoxically, substantially lower (38), reflecting stronger selection (weaker genetic drift) and lower mutation rates, respectively. Strong selection in salamanders might therefore "overwhelm" genetic drift in salamander species with low effective population sizes, contrary to expectations. High rates of selection might compensate for the low rates of mutation, a plausible explanation for the relatively low species richness in most salamander Family level clades: strong selection preserves slowly evolving salamander families from elimination by genetic drift.

In agreement with the proposal of selection for diversity, the MHC study also revealed a related correlation between dN and the amount of allelic variation. The authors concluded: "increased nucleotide substitution rate can promote allelic variation within lineages" (32). That conclusion again supports the proposal that a positional or regional effect on mutation rates can explain the correlation between dN and dS in terms of mutation/substitution balance, because positive selection is not expected to act on dS unless it is acting on *locus specific* mutation rates rather than on genes and codons.

A locus specific, regional explanation for the dN:dS correlation is consistent with these findings:

1) dN and dS in genes are correlated with polymorphisms in proximal UCEs, the majority of which (77 %) are located in intergenic or intronic sequences (35).

2) Balancing selection operates on different regions of the MHC, with the later replicating Class II region experiencing higher levels of polymorphism/mutation rates than either Class I or Class III (39).

Together, these and other observations suggest a role for a DNA replication-timing effect on the genome wide distribution of mutations in both coding and non-coding DNA: sub-chromosomal DNA replication timing domains, which are located non-randomly across the genome, expose coding and non-coding DNA simultaneously to the same sources of mutation.

3. DNA Replication Timing (RT)

3.1 RT and Replication Origins.

The eukaryotic genome is broadly partitioned into two spatial and temporal compartments: early replicating (open) euchromatin (EC) and late replicating (compact) heterochromatin (HC). EC is enriched in GC nucleotides while constitutive HC and facultative HC are enriched in AT nucleotides. In all species so far examined, mutation rates are significantly higher in genome regions containing constitutive or facultative heterochromatin compared to regions containing euchromatin (40, 41, 42, 43, 44).

The eukaryotic DNA replication-timing program has been intensively studied over the last several years in yeast and metazoa (45). Briefly, the RT program corresponds to replication domains, or extensive regions of several hundred kilobases to several megabases of subchromosomal DNA in vertebrates and 75 to 250 kb regions in invertebrates (46). Hence replication domain size scales with genome size. Each domain is replicated at a different time during S phase: early S, mid S and late S phases (Figure 3). Replication of domains therefore follows a strict spatio-temporal program. Replication origins (start sites of DNA synthesis) are activated stochastically but homogenously within domains, and the domains themselves are replicated asynchronously during the cell cycle (47, 48).

In most species, the genetic locations of replication origins are not specified by a conserved DNA sequence. Instead, origin locations depend primarily on chromatin context. Although all origins are "licensed" by an origin recognition complex (ORC and MCM helicases) in late mitotic M phase and post divisional G1 phase, only about 10% of licensed origins are activated during S phase (49). Activation, or initiation, of replication origins occurs stochastically and with increasing density (initiations per kilobase) into mid-S phase (Figure 3), and then decreases as the cells progress toward the G2 phase prior to mitosis (M phase; 50). The replication domains, in contrast, follow a regular timing program as S phase progresses (Figure 3B).

A

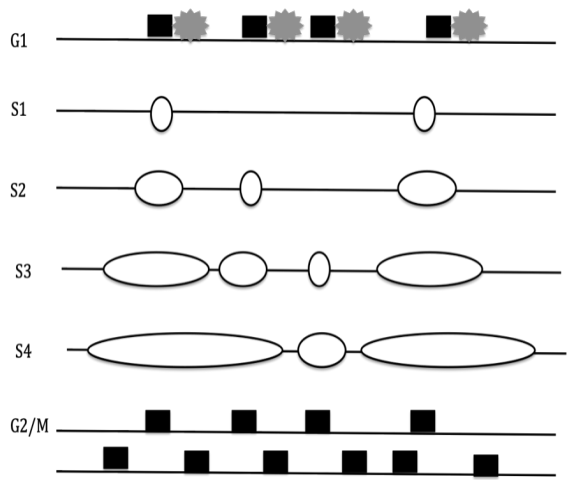

B

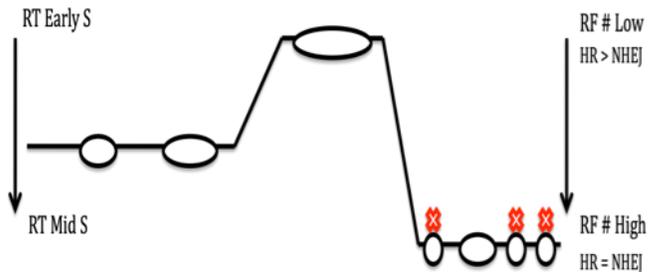

**Figure 3. A)** Schematic representation of the replication-timing program (RT). Black boxes: origin recognition complex (ORC). Gray serrated ovals: MCM helicases. Each stage of the cell cycle is represented (G1, S, G2/M). Note that replication fork density increases to mid S-phase (S2 – S3) before decreasing in late S-phase (S4). The figure depicts the binding of ORC during the G2/M phases of the cell cycle. The chromatin bound ORC recruits the MCM helicase complex in G1 of the following cell division cycle. The fully licensed origins in G1 then recruit a complex of initiation factors that assemble the active replication forks that define S phase. Note that origins do not fire synchronously within a given RT domain (depicted here), but fire homogenously to maintain a minimum origin-to-origin distance. **B)** Schematic Representation of Replication Timing Domains. Circles: potential replication origins. Red "x": checkpoint inhibited replication origins. Ovals: replication forks. The size of the oval corresponds to the timing of origin firing during S phase (fork rates are assumed constant within a given domain). Three replication-timing domains are depicted, extending from early to mid-S phase. RF #: replication fork density. At mid-S phase replication fork densities increase to a maximum before decreasing after mid-S phase. HR: allelic homologous DNA recombination repair. NHEJ: error prone non-homologous end joining DNA repair. Slanted lines: RT switch regions (see: 39, 138).

The timing of origin activation depends on many factors that are associated with two fundamental features:

1) The efficiency, or probability, of origin activation, which is determined by the number of MCM helicases loaded at the origin (51), and in budding yeast the levels of a nuclear complex of six positively acting initiation factors.

2) The strength of the intra-S phase checkpoint, which is a complex of factors that negatively regulates (represses) origin activation.

Other factors, such as Rif1 in higher eukaryotes and sir2 and rpd3 in yeast (52, 53), also play a role in establishing a late replication regime. While the biochemistry of these processes is beyond the scope of this paper, it is, in part, the opposing interplay between origin efficiency and the strength of origin inhibition that coordinates origin activation with gene transcription and establishes a late replication-timing program.

It might be interesting to note that origins located near or in highly expressed genes, which are early replicating, are more prone to DNA damage, while at the same time early replicating regions have substantially lower rates of substitution (54, 55). One possible explanation might be that origin-associated DNA damage, in addition to a relatively higher density of single strand DNA, up-regulates the intra-S checkpoint, which in turn represses later activated/less efficient origins and stimulates the DNA damage detection and repair system (DDR). Enhanced DDR function then targets earlier replicating DNA to repair more efficiently the DNA in the region where the origins reside. Early activated origins thus promote DNA repair in early replicating regions while repressing replication origins in later replicating regions.

3.2 RT and Genome Stability

Based on these and other findings, it has been suggested that the replication timing program evolved to govern and regulate the transcriptome during development, and to obviate the mutagenic effects of synchronous origin firing and a surfeit of multiple, simultaneously elongating DNA replication forks, which can be mutagenic and result in genome instability (56, 57, 58). Ablating the checkpoint, which inhibits late firing origins, causes origins to fire earlier and results in massive DNA damage (56). Conversely, over-expressing initiation-specific proteins causes all origins to fire earlier in budding yeast, and is lethal unless ribonucleotide reductase is simultaneously over-produced to supply the forks with sufficient levels of dNTPs (59).

Another related explanation for the evolution of replication timing therefore concerns the intra-S checkpoint and the associated DNA damage detection and repair system (DDR). This feature relates to dNTP supply and to the fact that replication forks are sensitive to dNTP levels, which, when imbalanced or perturbed, are highly mutagenic and a major cause of genome instability (rearrangements, amplifications and deletions, etc.). Hence, replication fork rates determine origin usage under replication stress in all organisms including bacteria. So-called "dormant", or auxiliary, origins in eukaryotes are activated in response to perturbed or stalled replication forks, and reduce the distances between replication origins to maintain the time required to duplicate the corresponding replication domain (60, 61).

Origin usage and replication fork rates are therefore universally correlated (62). This is likely to be the case even under non-stressed conditions (63, 64), and hence arguably replication fork rates, widely varying across the genome, coordinate the replication timing program (65): larger replicons (origin to origin distances) correlate with faster replication fork rates while smaller replicons correlate with slower fork rates. In this manner, the RT program maintains a constant overall rate of replication in replication domains of differing size, an essential feature of genome stability. At the same time, coordination of fork rates and origin efficiencies serves to limit the rate of DNA damage and maintain mutation/substitution balance across the genome.

3.3 RT and DNA repair

The DDR employs two principal systems that respond to and repair lethal DNA double strand breaks (DSBs): error free homologous recombination (HR) and error prone non-homologous end joining (NHEJ). Allelic HR depends on a homologous sister chromatid to repair DSBs and is largely restricted to S phase and predominates primarily in S and G2 phases (66, 67). NHEJ operates throughout the cell cycle and progressively replaces HR in the last half of S-G2 phase (68, 69, 70). This might explain why late replicating DNA has relatively higher mutation rates than early replicating DNA, although other factors such as error prone DNA damage polymerases play important roles (40, 71).

The relative ratios of these two repair systems thus directly impact mutation rates across the genome; and not surprisingly in a genome size dependent manner, with the faster acting NHEJ being three times more efficient than HR despite being more error-prone (66, 70). Eukaryotes with small genomes such as yeast rely predominantly on homologous recombination, while species with larger genomes such as vertebrates increasingly rely on NHEJ as C-values (haploid genome size in picograms) increase in the respective species (72, 73). Moreover, intron densities increase in larger, more NHEJ dependent genomes (73). Mutation rates are therefore expected to be anti-correlated between early and late replicating DNA in a genome size dependent manner: a weaker anti-correlation in small genome species; a stronger anti-correlation in large genome species.

Consequently, the replication-timing program becomes more deterministic as genome size increases (74, 75). In yeast, for example, although origins are initiated throughout S-phase (early to late), the vast majority of origins initiate in the first third of S phase (46), with the pattern of origin activation during S phase varying considerably from cell to cell (76). Species with larger genomes, in contrast, have significantly less flexible RT programs (46, 74). In vertebrates, late firing origins rarely if ever fire early in S phase, although the RT program itself is subject to species and tissue dependent differences in replication timing. At least half of the mammalian genome, for example, contains developmentally regulated replication domains that change replication timing from earlier to later replication and vice verse (77, 78).
.
3.4 RT and Genome Evolution

Although the replication-timing program becomes more complex with increasing genome size (C-value), the RT program itself is subject to selection and evolution, as well as to variation in replication timing in facultative HC (fHC) during tissue differentiation and development (78, 79). Most of the RT variants are associated with weak, late firing origins, which have a greater probability of loss during evolution (79, 80, 81). In addition to gene and karyotype evolution, the RT program likewise diverges widely between different lineages, yet DNA replication timing is largely conserved between closely related species (79, 80, 81).

Because evolution of replication-timing programs aligns with phylogeny, evolutionary changes in RT represent a third class, or source, of molecular evolution and speciation that recapitulates the phylogenetic tree, for example in primates and yeast (79, 80, 81, 82). Mutagenic loci (sites of higher sequence divergence) such as the human accelerated region (HAR) were found to be biased toward late replication, while sites of highly conserved sequences such as ultra-conserved elements and loss of function intolerant genes replicate early; divergent sites and HARs are enriched in replication timing variant regions, which are genomic regions that have experienced an evolutionary change in replication timing (83).

The evolution of the replication-timing program might provide a solution of sorts to the puzzling question of why the eukaryote genome has retained, rather than eliminated, so much, and such a variety of, non-coding and potentially maladaptive "junk" DNA. The evolution of the program involves the re-organization of the genome into differential compartments of facultative heterochromatin, constitutive heterochromatin and euchromatin. Evolutionary changes in chromatin structure and genome architecture represent adaptations to generate and maintain gene polymorphisms and allelic diversity in faster evolving later replicating genes. Many of these genes are "adaptive" and interface with the environment, such as tissue specific genes and the olfactory gene receptor complex (32, 34, 39).

4.1 Adaptation Cycles: Population Bottle Necks, Genetic Drift and Selection

The commonly held view of relaxed selection occurring in species with low effective population size, such as salamanders, is increasingly in doubt (38, 84, 85, 86). This came as something of a surprise since genetic drift can explain karyotype diversity among salamander phylogenetic clades; but it does not explain genetic diversity in salamander genes (87, 88), which varies little among clades and is more subject to selection compared to other vertebrates ($dN/dS > 1$).

The hypothesis that small effective population sizes and genetic drift explain changes in genome architecture and species richness is nonetheless comforted by the observation that speciation events have been found to be associated with higher substitution rates (17, 89, 90). Relaxed selection in small effective and census population sizes, accordingly, results in respectively higher rates of mutation in a context of lower genetic diversity. Consequently, the higher mutation rate would, as the population expands, result in subsequently higher levels of the standing genetic diversity on which positive and purifying selection can act (91, 92) This hypothetical scenario suggests that repeated cycles of drift during population bottlenecks followed by selection act synergistically to drive speciation and rates of species accumulation in phylogenetic clades (93, 94; Figure 2)

5. Maximum Lifespan and Peto's Paradox

5.1 The DDR and Variation in Body Mass: a Possible Link?

Darwin's "abominable mystery" addressed the geologically recent angiosperm radiation, considered the largest radiation in the terrestrial Tree of Life (95). The topology of the angiosperm phylogenetic tree resembles that of other lineages with highly imbalanced taxonomic clades—similar to mammals and salamanders—in terms of karyotype diversity, species richness, species evenness, and, additionally, range of C-value (96, 97; see Figure 1).

The observation of a correlation between dN and dS—whether or not a species (or region of the genome) is undergoing genetic drift or ecological selection—provides striking evidence for a positional effect influencing mutation rates associated with the replication-timing program: highly expressed, early replicating genes are selected for correct protein folding under a regime of purifying selection (72), while later replicating and tissue specific/developmental genes are selected for allelic diversity and phenotypic diversification (differentiation and speciation) under a regime of balancing selection.

Mutation/substitution rates vary between folding-selected genes (house keeping) and function-selected genes (adaptation and development) according to the relative strength of the checkpoint and effectiveness of the DDR in the respective genomic regions, or replication domains. The strength of the intra-S checkpoint can therefore be measured in terms of genome size (C-value), because more origins necessitate stronger checkpoint inhibition of the more numerous late activated origins in order to prevent them from competing for dNTPs with earlier firing origins (98, 99, 100, 101).

TE-driven genome expansion, for example, would lead to a greater probability of fork stalling and DNA damage and therefore would become maladaptive beyond a threshold C-value and whole body DNA content (102). Large genomes and large body sizes ultimately become maladaptive for opposing reasons. Large genomes necessitate stronger checkpoint and DDR functions entailing increasingly lower mutation/substitution rates, lower adaptive potential and hence higher extinction risk. Larger body sizes should, however, result in more DNA damage and higher lethal mutation rates, lower adaptive potential and hence a higher risk of extinction. The latter is known as Peto's paradox.

Maximum life span (MLS) can plausibly resolve Peto's paradox, and potentially serve as a proxy variable to measure the relative effectiveness of the DDR. Two observations support that proposal:

1) Peto's paradox addresses the observation that MLS strongly correlates with body mass (mammal Order level clades: 0.80; *p = 6 x $10^{-4}$; Family level clades: adjusted $R^2$ = 0.73; *p = 2 x $10^{-16}$; Table 1); yet, unexpectedly, long-lived large body mammals, having more cells and therefore cell division cycles, are significantly less prone to cancer and other mutation-associated disease (103, 104; Figure 5), an observation that is more related to cell cycle/cell size homeostasis than bulk body size (105, 106, 107, 108, 109, 110, 111).

2) Long-lived small body species, such as the Naked Mole Rat, have enhanced NHEJ and other DDR systems, but a normal mammalian C-value/cell size of about 3 pg (112, 113).

The proposal made here (MLS is a plausible proxy for the DDR—see: 112, 113—and C-value is a plausible proxy for the intra-S checkpoint) is at best tentative given that maximum life span and C-value are not themselves correlated (adjusted $R^2$ = 0,007, *p = 0.5; Table 1). Resistance to cancer causing radiation treatment, however, correlates with an up-regulated DDR (114), indicating that more effective DNA double strand break repair contributes not only to radiation tolerance but also to increasing maximum life spans.

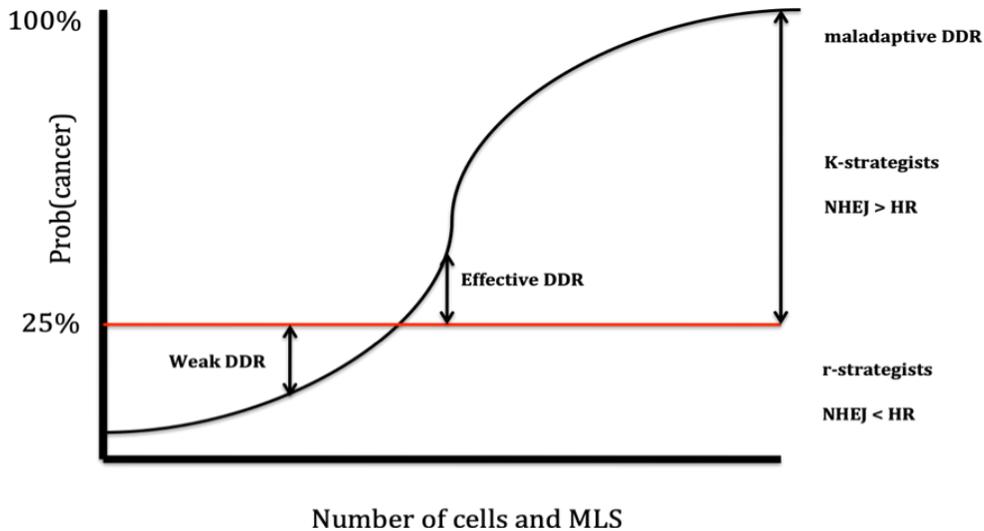

**Figure 4.** Illustration of Peto's paradox (adapted from 104). Y-axis: cancer risk. X-axis: body mass. Red line: observed cancer risk. Black curve: theoretically expected cancer risk. Black arrows reflect different strengths of the DDR, from weak DDRs that are maladaptive (small observed body size; higher than expected cancer risk) to strong DDRs that are likewise maladaptive (large observed body size and lower than expected cancer risk, but elevated extinction risk). The intersection of the black and red curves indicates the threshold at which DDR efficiency switches from favoring an unstable karyotype (r-strategist) to favoring an increasingly stable karyotype (K-strategist). Expanding genome sizes (higher TE density, CpG density and HC content; see references 134, 135) might therefore be fortuitous TE-driven adaptations that enhance the DDRs and, consequently, promote increases in body mass (see references 136 and 137). Transposable elements, rather than purely selfish parasites, act instead as genome commensals and mutualists benefiting from—and proliferating as a result of—the new niches and ecological adaptations TEs made possible for their respective hosts (Cope's rule: selection for K-strategists with larger genome size, body mass and longer maximum lifespan).

5.3 Genome Stability and Life History Traits

Although there is no a priori reason to expect a relationship between C-value and maximum life span, a yet to be established direct correlation might exist between the DDR and MLS, as well as a correlation between the intra-S checkpoint strength and C-value. The question of interest here concerns what specific molecular components of either system are potentially implicated (and how might they be implicated) in the established positive correlation between body mass and MLS (mammal Order level: adjusted $R^2 = 8.0$; *$p = 0.0006$), and the negative correlation between body mass and species richness (mammal Order level: adjusted $R^2 = 0.56$; *$p = 0.01$); and how might the relationships scale with each other, eg. linearly, geometrically or as a power law?

Moreover, gestation time has a significant negative relationship with neoplasia and malignancy prevalence, while at the same time neoplasia prevalence and somatic mutation rates are closely associated: species with fewer somatic mutations exhibit lower levels of neoplasia (109). It is well known, for example, that gestation time (which is related to body size) scales with embryonic growth rate (Table 2), suggesting slower rates of cell growth and division (115, 116, 117). It seems reasonable then to assume that a longer S phase and slower cell cycle would allow more time to repair DNA lesions, and hence serve to enhance genetic integrity and genome stability.

| Table 2 Species | Gestation time | Embryo Cell Cycle Duration |
|---|---|---|
| Drosophila | (24 hours) | 8 – 10 minutes |
| Frog | (6 – 21 days) | 0.5 hours |
| Salamanders | (14) – 728 days | 4 – 8 hours |
| Mouse | 19 – 21 days | 2 – 4 hours |
| Rabbit | 30 – 32 days | 5 – 8 hours |
| Dog | 58 – 69 days | 8 – 12 hours |
| Naked mole rat | 66 -77 days | NA |
| Beaver | 105 - 107 days | NA |
| Human | 280 days | 12 – 24 hours |
| Cow | 279-292 days | 32 hours |
| Elephant | 660 days | 18 – 36 hours |

**Table 2.** Association between gestation time and early embryo cell cycle duration. The parentheses indicate oviparous reproduction.

Of equal interest is the positive correlation in mammals between maximum life span and synteny conservation (adjusted $R^2 = 0.48$; *$p = 0.03$). The conservation of synteny blocks over 180 million years of karyotype evolution in mammals (118), for example, is a clear indicator of selection acting on genome stability via physiological and adaptive functions (macro-evolution impacting micro-evolution). The significant correlation with MLS, however, suggests that conservation of synteny blocks is also a feature of increased genome stability and a more effective DDR: a stable genotype imbedded in a stable karyotype that is, nevertheless, evolving much faster than the corresponding genotype.

Notably, synteny conservation does not associate significantly with species richness ($R^2 = 0.18$; *$p = 0.1$), whereas MLS, in contrast, is significantly associated (negatively) with SR (Order level mammals: adjusted $R^2 = 0.59$; *$p = 0.016$). Taken together, these observations suggest a role for the DDR—if MLS does in fact serve as a proxy for the DDR—in enhancing genome stability and in constraining rates of speciation and therefore levels of species richness. It would appear then that evolutionary changes at the sub-cellular level (DDR) promote evolutionary changes at the level of the organism (maximum life span and body mass) and at the level of phylogenetic clades (species richness). This hypothesis, however, warrants further investigation.

6. Conclusion

6.1 Rates of Karyotype Evolution Correlate with Species Richness in Mammals

At the taxonomic family level in mammals, species richness is known to correlate significantly with rates of karyotype rearrangements (genome scale changes: adjusted $R^2 = 0.42$; *$p = 3.3 \times 10^{-10}$), but not with rates of sub-chromosomal changes (adjusted $R^2 = 0.07$; *$p = 0.06$). The hypothesis that the imbalance in the Mammalian phylogenetic tree is due to the association between species richness and karyotype diversity—and therefore can be attributed to submicroscopic factors, presumably cellular and nuclear in origin—might apply also to angiosperms and all other metazoans (119, 120, 121).

It has been argued here, however, that the "submicroscopic factors" that account for the karyotype diversity-species richness correlation correspond to the close coordination between the replication-timing program, the transcription program, mutation rates and the DDR, with the related interplay between genome stability and instability (mutation/substitution balance) accounting, at least in part, for the dN-dS correlation and karyotype diversification. Although ecologically and molecularly independent, micro-evolutionary and macro-evolutionary processes likely intersect in mutually establishing speciation rates and species richness (122, 123). Changes in genome architecture occurring in parallel to and in conjunction with changes in life history traits (K versus r-selected species) might account, at least in part, for species richness and species evenness in the different vertebrates lineages (Figure 5).

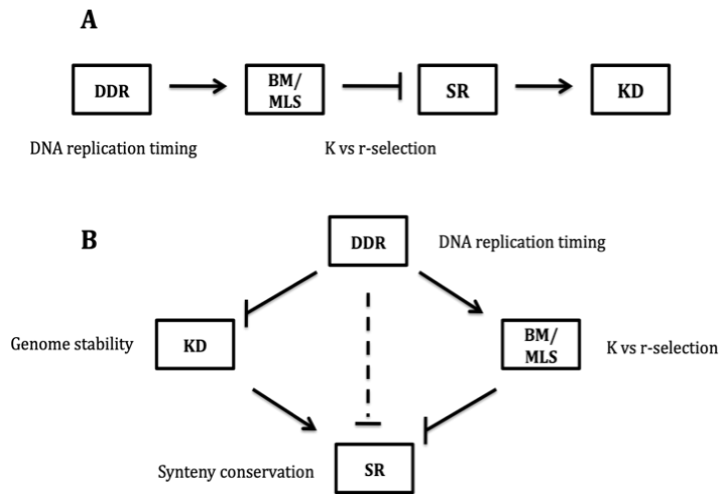

**Figure 5**: Direct (**A**: Darwinian selection) and Parallel (**B**: genetic drift couple to selection) Hypothetical Pathways Explaining the Karyotype-Species Richness Correlation in Mammals. DDR: DNA Damage detection and Repair systems; KD: Karyotype Diversity; BM: Body size; MLS: Maximum Life Span; SR:

Species Richness. Arrows and T-bars indicate positive and negative correlations, respectively. Dashed T-bar indicates an indirect negative effect. The correlations in Table 1 suggest alternative scenarios linking DNA repair mechanisms to species richness. **A.** Darwinian selection and adaptive radiation (Scenario 1): selection (K vs. r-selection, for example) operating on body size (BM) and maximum life span (MLS) explains the variation in species richness and species evenness among lineages and subclades within lineages (Class, Order, Family). This scenario assumes that the DDR has negligible effects on either lineage or subclade diversity. Macro-evolutionary forces acting on BM and MLS predominantly explain the correlations in Table 1: karyotype diversity (KD) is therefore a trivial consequence of clade-specific species richness. Body size and maximum life span are strongly correlated independently of phylogenetic crown age (Order level: adjusted $R^2 = 0,80$; *$p = 6 \times 10^{-4}$; Family level: $R^2 = 0.73$; *$p = 2 \times 10^{-16}$), suggesting that similar or related physiological processes account for the strong lineage specific correlation. The other correlations in Table 1 break down at the Family taxonomic levels in mammals (more recent phylogenetic clades), except for the correlation between SR and KD (Family level adjusted $R^2 = 0.43$; *$p = 3.3 \times 10^{-10}$). The absence of a correlation between MLS and KD supports a scenario in which selection explains most of the KD-SR variation. It should be noted, however, that synteny conservation correlates with MLS but not with BM, indicating that the physiological processes relating MLS to BM, although apparently cooperative, are not identical. BM and MLS in the Naked Mole Rat, for example, are not correlated. **B.** Genetic drift and non-adaptive radiation (Scenario 2): Macro (organism) and micro-evolutionary (molecular) forces acting in conjunction and in parallel explain the variation in the relationship between SR and KD. Scenario 2 is consistent with Bengtsson's hypothesis: "the cytological factors of importance for karyotype evolution are all of a sub-microscopic nature." The different stages represented here (Boxes) correspond to Kimura's four-stage model of non-adaptive radiation (see Figure 2). Bengtsson's hypothesis would seem to exclude an ecological role in explaining the KD-SR correlation. Darwinian selection, however, acts on standing genetic and genomic diversity, and is implicit in the hypothesis: genetic diversity, on which selection acts, is generated by the DDR (see 124). Hence, Bengtsson's hypothesis is consistent with Kimura's molecular theory of non-adaptive radiation, and with the parallel (but uncoupled) fixation during radiation of karyotype diversity and species richness in a phylogenetic clade: genetic drift is indirectly and intermittently coupled and uncoupled to Darwinian selection during the descent/evolution of a given lineage, generating "adaptation cycles". In contrast to Scenario 1, Scenario 2 incorporates Bengtsson's hypothesis and Kimura's theory, and therefore the relative strength of the DDR might play a significant role in explaining the phylogenetic relationship between karyotype diversity and species richness (see 124).

If that proposal is neither unreasonable nor particularly novel, a corollary nonetheless would be that extremely low mutation rates in the later replicating speciation/adaptation related genes are maladaptive in those species having either an overly efficient checkpoint (large C-value) or an overly effective DDR (large NHEJ/HR ratio). A stronger DNA damage detection and repair system could be one factor that explains the correspondingly low species richness in those lineages, presumably due to high extinction/low adaptation rates.

Simply stated, hyperactive checkpoints and/or DDRs result in a long term elevated lineage specific extinction risk (in contrast to a short term clade specific risk; see 124) due to a correspondingly low mutation/substitution supply within the lineage, and consequently a low standing level of genetic and allelic diversity and smaller effective population sizes in subclades. If so, an extremely low mutation/substitution rate (asymptotically approaching zero), with correspondingly low adaptive potential and elevated extinction risk, would likewise impose a ceiling on the evolution of genome sizes, body sizes and their related life history traits (Figure 4).

A causal relationship between genome/karyotype stability, maximum lifespan and cancer prevalence has yet to be firmly established, but the accumulating evidence is increasingly convincing (125, 126, 127). While substantial evidence supports a role for DNA repair systems in determining radiation resistance, maximum lifespan and other life history traits (K-strategists versus r-strategists), the roles DNA damage and repair potentially play in speciation can be outlined as follows:

    1) genome (in)stability drives genome evolution by either increasing or decreasing rates of karyotype evolution.
    2) genome evolution serves as a source of the standing genomic and allelic diversity on which selection can act.
    3) varying levels of genomic and allelic diversity result in varying degrees (magnitude) of adaptive radiation.

6.2 To summarize:

1) Karyotype diversity and species richness are correlated in mammals as a group.
2) The correlation between silent and non-silent mutation rates in eukaryotes can be explained by a genomic position effect on mutation rates.
3) DNA replication, damage and repair across broad regions of the genome (replication timing domains including coding and non-coding DNA) can account for the position effect: mutation rates uniformly increase during S phase.
4) The DDR mediates mutation/substitution balance across the genome.
5) Stronger DDRs can make possible larger body and longer living species (K vs. r-selection).
6) Selection acting on standing genomic and allelic diversity within isolated populations drives the correlation between karyotype diversity and species richness to fixation during an adaptive radiation.

Assessing the exact roles that gene and karyotype diversifications play in adaptive and non-adaptive radiations remains, however, an outstanding phylogenomic challenge (128, 129).